\documentclass[lettersize,journal]{IEEEtran}
\usepackage{xcolor,soul,framed} 

\usepackage{enumitem}
\colorlet{shadecolor}{yellow}
\usepackage[pdftex]{graphicx}
\graphicspath{{../pdf/}{../jpeg/}}
\DeclareGraphicsExtensions{.pdf,.jpeg,.png,.svg}
\usepackage[cmex10]{amsmath}
\usepackage{booktabs} 
\usepackage{array}
\usepackage{mdwmath}
\usepackage{mdwtab}
\usepackage{eqparbox}
\usepackage{url}
\usepackage{amssymb}
\usepackage{svg}
\usepackage{graphicx} 
\usepackage{cuted}
\usepackage{mathrsfs}
\usepackage{flushend}
\usepackage[utf8]{inputenc}
\usepackage[english]{babel}
\usepackage{amssymb,amsmath,amsthm}

\newcommand{\ie}{\textit{i}.\textit{e}.}
\usepackage{hyperref}
\hypersetup{
	colorlinks=true,
	linkcolor=blue,
	anchorcolor=blue,
	citecolor=blue}

\begin{document}

\title{NSI-IBP: A General Numerical Singular Integral Method via Integration by Parts}

\author{Shaolin~Liao$^*$,~\IEEEmembership{Senior Member,~IEEE}
\thanks{$^*$Shaolin Liao is with 1) School of Electronics and Information Technology (School of Microelectronics), Sun Yat-sen University, Guangzhou, Guangdong China 510006; 2) Department of Electrical and Computer Engineering, Illinois Institute of Technology, Chicago, IL USA 60616; and 3) Elmore Family School of Electrical and Computer Engineering, Purdue University,  West Lafayette, IN USA 47907.}

}




\maketitle

\begin{abstract}
A general framework of Numerical Singular Integrals (NSI) method based on the Integration By Parts (IBP) has been developed for  integrals involving singular and nearly singular integrands, or NSI-IBP. Through a general integration by parts formula and by choosing some analytically integrable function to approximate the original integrand, various well-known integration by parts methods can be derived. Rigorous mathematical derivations have been performed to transform the original singular or nearly singular integrals into non-singular integrals that can be computed efficiently, along with the boundary values added. What's more important, the NSI-IBP method works well even when the exact form of the singular integrand is not known. Criteria on how to choose the appropriate function with a known analytical integral that closely approximates the original integrand have been outlined and explained. Numerical recipe has been presented to apply the proposed NSI-IBP. Numerical experiments have been carried out on various singular integrals such as the power-law decaying integrand, the logarithmic function, and their hybrid products. It can be shown that various relative accuracy up to $10^{-15}$ can be achieved, even the exact singular function is not known. Finally, the nearly singular integrals involving the scalar Green's function have been evaluated for both electrostatics applications and Computational Electromagnetics (CEM) applications. 
\end{abstract}


\begin{figure}[th]
	\centering
	\includegraphics[width=0.5\textwidth]{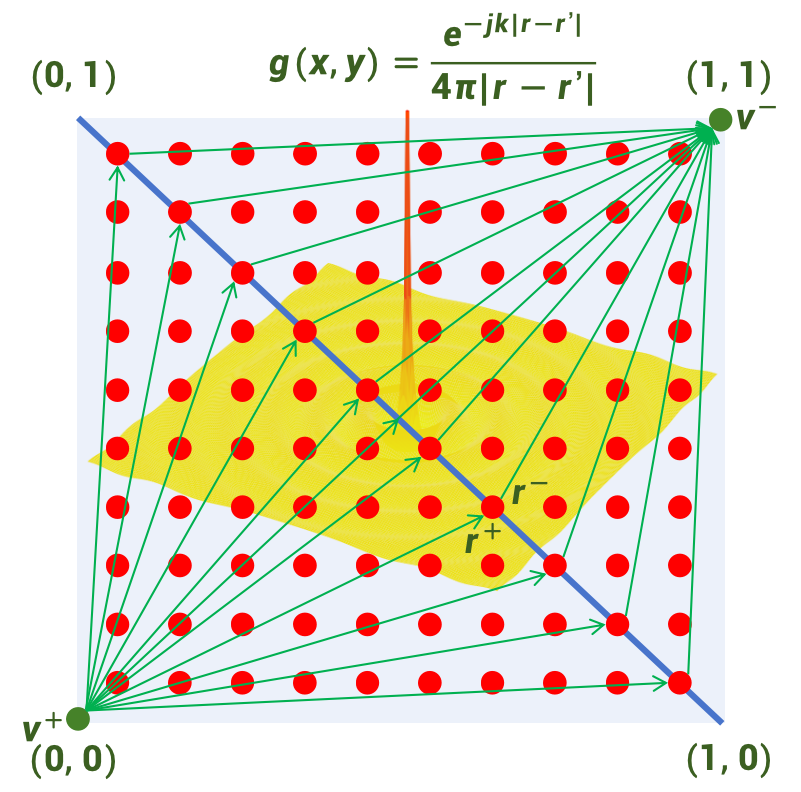}
	\caption{The convolution integral of scalar Green's function shown together with the RWG surface current on a pair of triangles. The red points denote the observation locations. Without loss of generality, the right triangles of unit length is used for the numerical experiment because any quadrilateral can be mapped to the unit square using bilinear transformation.}
	\label{fig:problem}
\end{figure}

\section{Introduction}
\IEEEPARstart{S}{ingular} integrals or nearly singular integrals exists in many physics problems and have attracted much attention in the past \cite{Li_2024}-\cite{Monegato_1994}, including electrostatic problems \cite{Khayat2005}-\cite{Wilton_1984} and electromagnetics problems \cite{Liao_Ping_Pong_2022}-\cite{Yaghjian_1980}. For example, one of the difficulty in Computational Electromagnetics (CEM) is the singularity due to the Green's function involved, which causes unavoidable singular integrals that need to be evaluated  \cite{Liao_Ping_Pong_2022},-\cite{Yaghjian_1980}. Similar problem happens in the Fourier spectral domain, where the Fourier spectrum of the green's function is singular  \cite{Liao_TAP_2009}.  

In some particular scenarios, analytical expressions exist for some particular singular or nearly singular integrals \cite{Tihon_2023}, \cite{Zuccotti_2023}, \cite{Tihon_2018}, \cite{Polimeridis_2010}. What's more, when the integrands of the singular or nearly singular integrals are known exactly, transform based methods can be employed \cite{Wang_2024}-\cite{Tang_2006}. However, Numerical Singular Integral methods (NSI) are much needed to carry out the general singular or nearly singular integrals where the exact forms of the integrands are not known, for example, when only numerical sampled values of the integrands are given, which is the focus of this paper.

As an introductory example in CEM, the scattered electric field at observation point $\overline{r}' = x' \hat{x} + y' \hat{y} + z' \hat{z} $ can be computed as the convolution integral of the scalar Green's function with the surface current, as shown in Fig. \ref{fig:problem} and can be expressed as follows \cite{Liao_Ping_Pong_2022} ,
\begin{flalign}\label{eqn:field_convolution}
	\frac{\overline{E}^s (\overline{r}')}{-j\omega \mu} &=  \overline{\overline{G}}_e (\overline{r}) \circledast \overline{J}(\overline{r})  \\
	&=\int\limits_{-\infty}^\infty \int\limits_{-\infty}^\infty \overline{\overline{G}}_e (x'-x, y'-y) \overline{J}(x, y) dx dy , \nonumber
\end{flalign}
where $\circledast$ denotes the 2D convolution operation; $\omega$ is the angular frequency; $\mu$ is the permeability; and the electric dyadic Green's function is given as Eq (\ref{eqn:dyadic_green}),
\begin{flalign}\label{eqn:dyadic_green}
	& \overline{\overline{G}}_e(\overline{r}) =   \left[\overline{\overline{I}} + \frac{1}{k^2} \nabla \nabla \right] g(\overline{r}),
\end{flalign}
where $\overline{\overline{I}}$ is the identity matrix; $k$ is the magnitude of the wave vector $\overline{k} = [k_x, k_y, k_z]$; and $g({\bf r}) $ is the scalar Green's function at the observation point vector $\overline{r} = [x, y, z]$,
\begin{flalign}
	g(r) = \frac{e^{-j k r }}{4 \pi r}, \ \ r = \left|\overline{r}\right|; \ \ k =|\overline{k}| = \omega \sqrt{\mu \epsilon}.\nonumber
\end{flalign} 

Similarly, the surface current can be expressed in terms of the convolution integral of the scalar Green's function and  the scattered electric field as follows \cite{Liao_Ping_Pong_2022}, 
\begin{flalign}\label{J_Es}
	&\overline{J}(x, y) =\overline{\overline{L}}_J (\overline{r}) \circledast \overline{E}^s(\overline{r})  \\
	=& \frac{4 \omega \epsilon}{j}  \left\{ \overline{\overline{I}} + \frac{1}{k^2} \begin{bmatrix}
		\frac{\partial^2}{\partial y^2} & -  \frac{\partial^2}{\partial x \partial y} \\
		-  \frac{\partial^2}{\partial x  y}   &   \frac{\partial^2}{\partial x^2} 
	\end{bmatrix} \right\}  \left[g(x, y) \circledast  \overline{E}_{//}^s(x, y)\right]. \nonumber
\end{flalign}


It is clear that both Eq. (\ref{eqn:field_convolution}) and Eq. (\ref{J_Es}) contains the following convolution integral involving the singular scalar Green's function,
\begin{gather}\label{conv_gJ}
	I(\overline{r}') \equiv g (\overline{r}) \circledast \overline{J}(\overline{r}) \\
	= \int_{c}^d \int_{a(y)}^{b(y)} \frac{e^{-j k \left[ (x'-x)^2 + R^2 \right]^{\frac{1}{2}}  }}{4 \pi \left[ (x'-x)^2 + R^2 \right]^{\frac{1}{2}} }  \overline{J}(x, y) dxdy , \nonumber
\end{gather}
with
\begin{gather}
	R = \left[ (y'-y)^2+(z'-z)^2 \right]^{\frac{1}{2}}.
\end{gather}

Without loss of generality, the RWG right triangle pair forming a square of unit length is used, as shown in Fig. \ref{fig:problem}. This is because any quadrilateral can be mapped to the square of unit length through bilinear transformation. Now the convolution integral of Eq. (\ref{conv_gJ}) for the positive right triangle reduces to the following,
\begin{gather}\label{conv_gJ2}
	I(\overline{r}')  = \int_{0}^1 \int_{0}^{1-y} \frac{e^{-j k \left[ (x'-x)^2 + R^2 \right]^{\frac{1}{2}}  }}{4 \pi \left[ (x'-x)^2 + R^2 \right]^{\frac{1}{2}} }  \overline{J}(x, y) dxdy ,  
\end{gather}
which can be further simplified  by change of variables $x' - x \rightarrow x$ and $y' - y \rightarrow y$,
\begin{gather}\label{singular_green}
	I_g(\overline{r}') \equiv \int_{0}^1 I_{\frac{1}{2}}(y) dy, 
\end{gather}
with
\begin{gather}\label{singular_green_1D}
	I_{\frac{1}{2}}(y) \equiv \int_{0}^{1-y}  \frac{ \overline{p}_0(x, y)}{ \left( x^2 + R_0^2 \right)^{\frac{1}{2}} }   dx, 
\end{gather}
where $R_0=\sqrt{ y^2+(z'-z)^2}$ and the following has been defined,
\begin{gather}
	\overline{p}_0(x, y) = \frac{e^{-j k \left[ x^2 + R_0^2 \right]^{\frac{1}{2}}  }}{4 \pi }  \overline{J}(x'-x, y'-y).
\end{gather}

It is clear that the integral $I_{\frac{1}{2}}(y)$ of Eq. (\ref{singular_green_1D}) is singular when $R_0 =0$ and nearly singular when $R_0 \neq 0$. So   efficient numerical singular or nearly singular integral method is much needed, which is the focus of this paper.

The main contributions of this article are given below,
\begin{enumerate}
	\item A general NSI framework based on a general IBP formula, \ie, the NSI-IBP method, has been presented and proved in a rigorous mathematical way;
	
	\item The NSI-IBP method effectively transform the original singular or nearly singular integral into the non-singular integral that can be computed numerically friendly, plus the boundary values;
	
	\item The NSI-IBP method works well for even when the exact singular or nearly singular integrands are not known, for example, when only numerical values at some given sample points of the integrands are given.
	
	\item The number of sample points for the NSI-IBP method can be optimized by appropriately chose functions that  approximate the singular or nearly singular integrands optimally to achieve the best computational cost and memory storage;
	
	\item Criteria on choosing the appropriate singular or nearly singular function that can best approximate the original singular or nearly singular integrand have been discussed; 
	
	\item Experiments on typical singular integrals such as the power-law singular integrals, the logarithmic singular integrals and their hybrid products, as well as nearly singular integrals met in the electrostatic and electromagnetics problems. The results show that the NSI-IBP method is very promising to compute various singular or nearly singular integrals, even when the exact form of the singular integrand is not known.
\end{enumerate}

Finally, for simplicity of presentation, only 1D case is shown, this is because, higher dimensions such as the 2D RWG integral of Eq. (\ref{singular_green}) can be reduced to many 1D integrals. 

\section{Numerical Singular Integral Methods}
It is clear that the convolution integrals involves singular integrands such as the scalar green's function, among others. Such singular integrals need to be regularized before they can be implemented numerically. 

Let's look at the following general singular integral,
\begin{flalign}\label{eqn:NSI_general}
	& I_0 = \int_a^b f(x)  dx = \int_a^b \frac{p(x)}{q(x)}  dx ,
\end{flalign}
where the integrand $f(x)$ is singular that can be generally expressed as follows
\begin{flalign}
	&   f(x) = \frac{p(x)}{q(x)},  
\end{flalign}
with $p(x)$ is assumed to be the polynomial of order $n$, \ie, $\mathcal{P}_n$, and $q(x)$ containing roots $x_k, k\ge 1$ within the range of integral $[a, b]$,
\begin{gather}
	p(x) \in \mathcal{P}_n; \ \ q(x_k) =0, \ \ k \ge 1.
\end{gather}

\section{Integration by Part Method}
The original integral $I_0$ can be computed through the integration by part method as follows,
\begin{flalign}\label{eqn:NSI_general_qx}
	& I_0 = \left. p(x) h(x)  \right|_a^b - \int_a^b  h(x) p'(x) dx ,
\end{flalign}
with 
\begin{gather}\label{Iq_part}
	h(x) = \int_a^b \frac{1}{q(x)} dx.
\end{gather}

However, the integration by part method requires $q(x)$ and the analytical formula of $h(x)$ are known, which sometimes is impossible.

\section{Background and Notations}\label{sec:Background}
In this Section, basic background knowledge on IBP is presented, along with definitions of notations.

\subsection{Simple Integration by Part Method}
One might try to transform the original integral of Eq. (\ref{eqn:NSI_general}) via the integration by part method as follows,
\begin{gather}\label{I0_by_part}
	I_0 =  \int_a^b \frac{1}{x^\gamma} \left[ x^\gamma f(x) \right] dx  \\
	=  \int_a^b  h_\gamma(x)' \left[ x^\gamma f(x) \right] dx, \nonumber  
\end{gather}
which can be then evaluated through the integration by part method,
\begin{gather}\label{I0_by_part_1}
	I_0 = \left. h_\gamma(x) \left[   f_\gamma(x) \right] \right|_a^b  - \int_a^b  h_\gamma(x) f_\gamma'(x)  dx,
\end{gather}
where
\begin{gather}
	f_\gamma(x) = x^\gamma f(x),
\end{gather}
and
\begin{gather} \label{int_qx}
	h_\gamma(x) \equiv \int \frac{1}{x^\gamma} dx = \left\{ \begin{matrix}
		\frac{x^{1-\gamma}}{1-\gamma}, \ \ \gamma \neq 1,   \\
		\ln(x), \ \ \gamma =1.
	\end{matrix}  \right., 
\end{gather}

Substituting Eq. (\ref{int_qx}) into Eq. (\ref{I0_by_part_1}), the following is obtained,
\begin{gather}\label{I0_by_part_2}
	I_0 = \left\{ \begin{matrix}
		\left. \frac{1}{1-\gamma} x f(x)   \right|_a^b  - \frac{1}{1-\gamma} \int_a^b  x^{1-\gamma} f_\gamma'(x)  dx,  \gamma \neq 1;  \\
		\left. x \ln(x) f(x) \right|_a^b  - \int_a^b  ln(x) \left[ f(x) + x f'(x) \right]   dx, \gamma =1.
	\end{matrix} \right.  
\end{gather}

In particular, when $\gamma=0$, Eq. (\ref{I0_by_part_2}) reduces to the following,
\begin{gather}\label{I0_gamma_1}
	I_0 = \left.   x f(x)   \right|_a^b  -  \int_a^b  x f_\gamma'(x)  dx \\
	= \left.   x f(x)   \right|_a^b  -  \int_a^b  \frac{x}{q(x)} \left[ p'(x)- \frac{p(x)}{q(x)} q'(x) \right]  dx.  \nonumber
\end{gather}

However, to compute the integral in Eq. (\ref{I0_by_part_2}) efficiently, the integrand at the singularity point $x_k$ should be finite,
\begin{gather}\label{qx_gamma0}
	\lim_{x\rightarrow x_k} \left\{ \left|h_\gamma(x) f_\gamma'(x) \right|\right\} = \lim_{x\rightarrow x_k} \left\{ \left| \frac{x^{1-\gamma}}{1-\gamma} f_\gamma'(x) \right|\right\} < \infty.
\end{gather}

Now assuming that $q(x) = x^{\tilde{\gamma}}$ in Eq. (\ref{qx_gamma0}) and the following is obtained,
\begin{gather}
	f_\gamma'(x) = \left( p(x) x^{\gamma -\tilde{\gamma}} \right)'  
	=  p'(x) x^{\gamma -\tilde{\gamma}} + p(x)  \left(  x^{\gamma -\tilde{\gamma}} \right)' \nonumber \\
	= \left\{ \begin{matrix}
		p'(x) x^{\gamma -\tilde{\gamma}} , \ \  \tilde{\gamma} = \gamma	; \\
		p'(x) x^{\gamma -\tilde{\gamma}} +   \frac{\left(\gamma -\tilde{\gamma} \right)  p(x) }{x^{1-\left(\gamma -\tilde{\gamma} \right) }},  \ \  \tilde{\gamma} \neq \gamma.
	\end{matrix} \right.
\end{gather} 
which means that $q(x)$ has to follow the power law decay of $x^\gamma$. So $\gamma$ needs to be known before Eq. (\ref{I0_by_part_1}) can be computed efficiently.  


\subsection{Fractional Integration by Part Method}
When $q(x)$ or the  analytical formula of $h(x)$ of Eq. (\ref{Iq_part}) is not known, a function $\tilde{q}(x)$ with known $\tilde{h}_q(x)$ can be chosen to approximately represent $q(x)$ and the original integral $I_0$ can be evaluated as follows,
\begin{gather}\label{tildeq}
	I_0  = \int_a^b \frac{1}{\tilde{q}(x)} \tilde{p}(x) dx ,
\end{gather}
with
\begin{gather}
	\tilde{p}(x) = \frac{\tilde{q}(x)}{q(x)} p(x) = \tilde{q}(x) f(x).
\end{gather}

Similar to Eq. (\ref{eqn:NSI_general_qx}), Eq. (\ref{tildeq}) can be evaluated through the integration by part method as follows,
\begin{flalign}\label{titlde_qx}
	& I_0 = \left. \tilde{p}(x) \tilde{h}_q(x)  \right|_a^b - I_1,
\end{flalign}
with 
\begin{gather}\label{Iq_part_tilde}
	I_1 = \int_a^b  \tilde{h}_q(x) \left[\tilde{q}(x) f(x)\right]' dx ,
\end{gather}
and
\begin{gather}\label{hs}
	\tilde{h}_q(x) = \int_a^b \frac{1}{\tilde{q}(x)} dx. 
\end{gather}


Since the integral $I_0$ in Eq. (\ref{eqn:NSI_general}) is singular, it is straightforward to reduces its value around its singularities by timing $f(x)$ with some suppressing weight function $\tilde{q}(x)$. So let's consider the following integral,
\begin{flalign}\label{eqn:partial_general}
	I_1 = \int_a^b   \tilde{h}(x)  \frac{ d }{d x} \left\{ \tilde{q}(x)  f(x)  \right\} dx.
\end{flalign}

The goal is to find some appropriate   functions $\tilde{q}(x)$ to meet the following criteria.

\begin{enumerate}
	\item Analytical form of $\tilde{h}(x)$ in Eq. (\ref{hs}) exists.
	
	\item The integrand of $I_1$ in Eq. (\ref{eqn:partial_general}) is finite and its derivatives  are continuous so that conventional integral methods such as the quadrature rules can be applied,
	\begin{gather}\label{condition1}
		I_1  = \int_a^b   \tilde{h}(x)  \frac{ d }{d x} \left\{ \tilde{q}(x)  f(x)  \right\} dx \\ 
		= \sum\limits_{j} w_j  \tilde{h}(x_j)  \frac{ d }{d x} \left\{ \tilde{q}(x_j)  f(x_j)  \right\}, \nonumber
	\end{gather}
	where $w_j$ is some quadrature weight \cite{Li_2024}, \cite{Lombardi_2024}, \cite{Chang_2021}-\cite{Kolm_2001}.
\end{enumerate}


\section{A Novel General Integration by Part Method}\label{sec:general}

In this Section, a general IBP method that can be used to derive all IBP variants in Section \ref{sec:Background} is presented.

Let's consider the following more general integral,
\begin{flalign}\label{eqn:partial_general0}
	I_1 = \int_a^b   h(x)  \frac{ d }{d x} \left\{ \tilde{q}(x)  f(x)  \right\} dx,
\end{flalign}
where $h(x)$ here might  not be the same as $\tilde{h}(x)$ as in Eq. (\ref{eqn:partial_general}).

\subsection{Unknown $q(x)$}
	

When $h(x) = \tilde{h}(x)$ , the original integral $I_0$ of Eq. (\ref{titlde_qx}) can be calculated as follows,
\begin{gather}\label{Io_hs}
	I_0 = \left. \tilde{h}(x)  \tilde{q}(x)  f(x)   \right|_a^b - \int_a^b   \tilde{h}(x)  \frac{ d }{d x} \left\{ \tilde{q}(x)  f(x)  \right\} dx  \\
	= \left. \frac{\tilde{h}(x)}{\tilde{h}'(x)}  f(x)   \right|_a^b - \int_a^b   \tilde{h}(x)  \frac{ d }{d x} \left\{  \frac{f(x)}{\tilde{h}'(x)}  \right\} dx, \nonumber
\end{gather}
which can be computed numerically as
\begin{gather}
	I_0 = \left. \frac{\tilde{h}(x)}{\tilde{h}'(x)}  f(x)   \right|_a^b - \sum\limits_{j} w_j  \tilde{h}(x_j)  \frac{ d }{d x} \left\{   \frac{f(x_j)}{\tilde{h}'(x_j)}  \right\},
\end{gather}
where Eq. (\ref{hs}) has been applied.

Now Eq. (\ref{Io_hs}) reduces to the following,
\begin{gather}\label{Io_hs_qx_app}
	I_0 = \tilde{I}_b - \tilde{I}_1, 
\end{gather}
with
\begin{gather}\label{I1_app}
	\tilde{I}_b \equiv \left. \tilde{h}(x)  \tilde{q}(x)  f(x)   \right|_a^b, \\
	\tilde{I}_1 \equiv  \int_a^b   \tilde{h}(x)  \frac{ d }{d x} \left\{ \tilde{q}(x)  f(x)  \right\} dx \nonumber \\
	= \int_a^b  \tilde{h}(x)  \tilde{q}(x) f'(x)  dx  + \int_a^b   \tilde{h}(x)  \tilde{q}'(x) f(x)  dx, \nonumber
\end{gather}
where $\tilde{h}'(x) = 1/\tilde{q}(x)$ has been used.

One possible choice of $\tilde{q}(x)$ is  $\tilde{q}(x) = x^{\tilde{\gamma}} $ and Eq. (\ref{I1_app}) reduces to 
\begin{gather}\label{I1_hsx_app_1}
	\tilde{I}_1 = \frac{1}{1-\tilde{\gamma}} \int_a^b x f'(x)  dx  + \frac{\tilde{\gamma}}{1-\tilde{\gamma}} I_0,
\end{gather}
where the following has been used,
\begin{gather}
	\tilde{q}'(x) = \tilde{\gamma} x^{\tilde{\gamma}-1}; \\
	\tilde{h}(x) = \frac{x^{1-\tilde{\gamma}}}{1-\tilde{\gamma}}; \nonumber \\
	\tilde{h}(x)  \tilde{q}(x) = \frac{x}{1-\tilde{\gamma}}; \nonumber \\
	\tilde{h}(x)  \tilde{q}'(x) = \frac{\tilde{\gamma}}{1-\tilde{\gamma}}. \nonumber
\end{gather}

Substituting Eq. (\ref{I1_hsx_app_1}) in to Eq. (\ref{Io_hs_qx_app}), the original integral $I_0$ is obtained as 
\begin{gather}\label{Io_hs_qx_app_1}
	I_0 = \left. x  f(x)   \right|_a^b -   \int_a^b x f'(x)  dx,
\end{gather}
which is exactly the result of the integration by part. 

It is clear Eq. (\ref{Io_hs_qx_app_1}) doesn't depend on $\tilde{\gamma}$, which means that the computational method of Eq. (\ref{Io_hs_qx_app}) doesn't depend on the choice of $\tilde{\gamma}$.

\subsection{Known $q(x)$}
When $q(x)$ is known, the integral $I_1$ of Eq. (\ref{eqn:partial_general}) can   be evaluated in different ways as follow.

\subsubsection{Approximate $\tilde{q}(x)$}
Eq. (\ref{Io_hs_qx_app}) reduces to the following,
\begin{gather}
	\tilde{I}_1 =  \int_a^b   \tilde{h}(x)  \frac{ d }{d x} \left\{ \frac{\tilde{q}(x)}{q(x)}  p(x)  \right\} dx  \\
	= \int_a^b   \tilde{h}(x)   \left\{ \left[ \frac{\tilde{q}(x)}{q(x)} \right]'  p(x)   + \frac{\tilde{q}(x)}{q(x)}  p'(x) \right\} dx  \nonumber \\
	= \tilde{I}_{11} + \tilde{I}_{12}, \nonumber  
\end{gather}
where the following have been defined,
\begin{gather}\label{I11_I12}
	\tilde{I}_{11} \equiv  \int_a^b \tilde{h}(x)  \frac{\tilde{q}(x)}{q(x)}  p'(x) dx, \\
	\tilde{I}_{12} \equiv \int_a^b   \tilde{h}(x)  \left[ \frac{\tilde{q}(x)}{q(x)} \right]'  p(x) dx = \tilde{I}_{121}  + \tilde{I}_{122}, \nonumber
\end{gather}

Now let's evaluate $\tilde{I}_{121}$ and $\tilde{I}_{122} $ in Eq. (\ref{I11_I12}) as follows.
\begin{enumerate}[label=(\arabic*)]
	\item Integral $\tilde{I}_{121}$: 
	\begin{gather}\label{I121}
		\tilde{I}_{121} \equiv  \int_a^b   \tilde{h}(x)  \tilde{q}'(x) \frac{p(x)}{q(x)} dx  \\
		=\int_a^b   \tilde{h}(x)  \tilde{q}'(x) f(x)  dx \nonumber   \\
		= \tilde{I}_b  - \int_a^b  \tilde{q}(x) \left\{ \tilde{h}'(x) f(x) + \tilde{h}(x) f'(x) \right\}    dx \nonumber   \\
		= - I_0 + \tilde{I}_b      - \int_a^b  \tilde{h}(x)  \tilde{q}(x) f'(x)  dx. \nonumber   
	\end{gather}
	
	\item Integral $\tilde{I}_{122}$:  
	\begin{gather}\label{I122}
		\tilde{I}_{122} \equiv - \int_a^b   \tilde{h}(x) \tilde{q}(x)   \frac{q'(x)}{q(x)} f(x)   dx \\
		= I_0 - \int_a^b   \left[\tilde{h}(x) q(x) \right]'   \frac{\tilde{q}(x) }{q(x)} f(x)   dx  \nonumber \\
		= I_0 - \tilde{I}_b   + \int_a^b  \tilde{h}(x) q(x) \left[ \frac{\tilde{q}(x) }{q(x)} f(x) \right]'  dx. \nonumber  
	\end{gather} 
	
\end{enumerate}

Substituting $I_{121}$ of Eq. (\ref{I121}) and $I_{122}$ of Eq. (\ref{I122}) into Eq. (\ref{I11_I12}), $\tilde{I}_{12}$ is obtained, 
\begin{gather}\label{I12_2}
	\tilde{I}_{12} = \int_a^b  \tilde{h}(x) \tilde{q}(x) \left\{ \frac{q(x)}{\tilde{q}(x)} \left[ \frac{\tilde{q}(x) }{q(x)} f(x) \right]' -f'(x) \right\} dx \\
	= \int_a^b  \left[\ln\left(\frac{\tilde{q}(x) }{q(x)}\right) \right]' \tilde{h}(x) \tilde{q}(x)  f(x)  dx \nonumber \\
	= \left. \ln\left(\frac{\tilde{q}(x) }{q(x)}\right)   \tilde{h}(x) \tilde{q}(x)  f(x)  \right|_a^b \nonumber \\
	-  \int_a^b  \ln\left(\frac{\tilde{q}(x) }{q(x)}\right)  \tilde{h}(x) \tilde{q}(x)  f'(x)  dx \nonumber \\
	-  \int_a^b  \ln\left(\frac{\tilde{q}(x) }{q(x)}\right) \left[\tilde{h}(x) \tilde{q}(x) \right]'  f(x)  dx. \nonumber  
\end{gather} 

Substituting Eq. (\ref{I12_2}) and Eq. (\ref{I11_I12}) into Eq. (\ref{Io_hs_qx_app}), the original integral $I_0$ is obtained as follows,
\begin{gather}\label{I0_all}
	I_0 = \tilde{I}_b - \left. \ln\left(\frac{\tilde{q}(x) }{q(x)}\right)   \tilde{h}(x) \tilde{q}(x)  f(x)  \right|_a^b\\
	- \int_a^b \tilde{h}(x)  \frac{\tilde{q}(x)}{q(x)}  p'(x) dx +  \int_a^b  \ln\left(\frac{\tilde{q}(x) }{q(x)}\right) \left[\tilde{h}(x) \tilde{q}(x)  f(x)  \right]'  dx. \nonumber 
\end{gather}


\subsubsection{Conventional   Integration by Part Method}


It can be shown that the integration by part method of Eq. (\ref{eqn:NSI_general_qx}) is a special case of Eq. (\ref{I0_all}) when $\tilde{q}(x) = q(x)$.

\subsubsection{Direct Calculation Methods} \label{sec:direct}
When $q(x)$ is known, the integral $I$ of Eq. (\ref{eqn:partial_general}) can also be evaluated directly as follows, 
\begin{gather}\label{eqn:partial_general_alt}
	I_1  =   I_{11} + I_{12} + I_{13},  
\end{gather}
with
\begin{gather}
	I_{11} \equiv \int_a^b   h(x) \frac{ d \tilde{q}(x)  }{d x} \frac{p(x)}{q(x)}   dx, \\
	I_{12} \equiv - \int_a^b   h(x) \tilde{q}(x) \frac{ d  q(x) }{   d x} \frac{   p(x)}{q^2(x)}  dx, \nonumber \\
	I_{13} \equiv   \int_a^b    \frac{h(x)\tilde{q}(x)}{q(x)}  \frac{ d p(x)  }{d x}  dx, \nonumber  
\end{gather}

By setting Eq. (\ref{eqn:partial_general}) equal to Eq. (\ref{eqn:partial_general_alt}), the following is obtained,
\begin{gather}\label{eqn:partial_general_both}
	I_{+} = \left. h(x)  \tilde{q}(x)  \frac{p(x)}{q(x)}    \right|_a^b  - I_{13},
\end{gather}
with
\begin{gather}\label{I12}
	I_{+} \equiv \int_a^b   \frac{d h(x)}{d x}  \tilde{q}(x)  \frac{p(x)}{q(x)} dx + I_{11} + I_{12} \\
	= \int_a^b  \left\{ \frac{d }{d x} \left[ h(x)  \tilde{q}(x)  \right] -   \frac{h(x)  \tilde{q}(x)}{q(x)} \frac{ d  q(x) }{d x} \right\} \frac{p(x)}{q(x)}    dx,   \nonumber
\end{gather}

Now let's look at different situations as follow.

To recover the original integral $I_0$, one of the following conditions should be satisfied.

\begin{enumerate}[label=(\arabic*)]
	\item $h(x) \tilde{q}(x) = x$: 
	when $h(x) \tilde{q}(x) = x$, Eq. (\ref{I12}) reduces to
	\begin{gather}\label{I12_hw}
		I_{+} = I_0 -  \int_a^b  \frac{x}{q(x)} \frac{ d  q(x) }{d x} \frac{p(x)}{q(x)}    dx,   \nonumber
	\end{gather}
	and the original integral $I_0$ is obtained as
	\begin{gather}\label{hsx}
		I_0 -  \int_a^b  \frac{x}{q(x)} \frac{ d  q(x) }{d x} \frac{p(x)}{q(x)}    dx = \left. x \frac{p(x)}{q(x)}    \right|_a^b \\
		- \int_a^b    \frac{x}{q(x)} \frac{ d p(x)  }{d x} dx, \nonumber  
	\end{gather}
	which doesn't depend on the choice of $\tilde{q}(x)$ and it is actually the result given in Eq. (\ref{I0_gamma_1}) by the integration by part method with $\gamma =0$.
	
	\item $h(w) \tilde{q}(x) q'(x)/q(x) = -1$: Eq. (\ref{I12}) reduces to
	\begin{gather}\label{I12_hwdq}
		I_{+} = I_0 -  \int_a^b  \frac{d }{d x} \left[ \frac{q(x)}{q'(x)}  \right]  \frac{p(x)}{q(x)}    dx,   \\
		=   -  \int_a^b \frac{q''(x)}{\left[q'(x)\right]^2} p(x)   dx , \nonumber
	\end{gather}
	where $q'(x)$ and $q''(x)$ denote the first derivative and second derivative respectively.
	
	In order to recover the original integral $I_0$, $q(x)$ should also satisfies the following relation,
	\begin{gather}\label{I12_hwdq2}
		\frac{q''(x)}{\left[q'(x)\right]^2} = - \frac{\beta}{q(x)},
	\end{gather}
	or equivalently,
	\begin{gather}\label{I12_hwdq3}
		\left[ \ln q'(x) \right]'= -\beta \left[ \ln q(x)  \right]',  \\
		\ln q'(x) = -\beta  \ln q(x) + c_0, \nonumber \\ 
		q'(x) =  c q^{-\beta}(x),  \nonumber
	\end{gather}
	from which Eq. (\ref{I12_hwdq}) reduces to the following,
	\begin{gather}\label{I12_hwdq_1}
		I_{+} =  \beta I_0. \nonumber
	\end{gather}
	
	Some possible $q(x)$ that satisfies Eq. (\ref{I12_hwdq3}) are given below.
	
	\begin{enumerate}
		\item Power-law decay:
		\begin{gather}
			\frac{1}{q(x)} = x^{-\gamma} \rightarrow \beta= \frac{1 - \gamma}{\gamma}, \ \ c = \gamma.
		\end{gather}
		
		\item Exponential decay:
		\begin{gather}
			\frac{1}{q(x)} = \exp(-\lambda x) \rightarrow \beta= -1, \ \ c = \lambda.
		\end{gather}

	\end{enumerate}

	\item $I_{+} = I_0$: to achieve $I_{+} = I_0$, the following condition should be met,
	\begin{gather}\label{condition}
		\frac{d u (x) }{d x}   - u(x)   \frac{1}{q(x)} \frac{ d  q(x) }{d x} =1,
	\end{gather} 
	where $u(x) \equiv h(x)  \tilde{q}(x)$ and Eq. (\ref{eqn:partial_general_both}) reduces to 
	\begin{gather}\label{eqn:partial_general_both2}
		I_0 = \left. u(x)  \frac{p(x)}{q(x)}    \right|_a^b  - \int_a^b    \frac{u(x)}{q(x)}  \frac{ d p(x)  }{d x}  dx.
	\end{gather}
	
	Eq. (\ref{condition}) is the ordinary differential equation of order one that can be readily solved as follows,
	\begin{gather}\label{solution}
		v(x) = \exp\left\{ - \int \frac{1}{q(x)} \frac{ d  q(x) }{d x} dx \right\} =  \frac{1}{q(x)}, \\
		u(x) = \frac{1}{v(x)} \int v(x) dx = h(x) q(x),
	\end{gather} 
	from which Eq. (\ref{eqn:partial_general_both2}) reduces to 
	\begin{gather}
		I_0 = \left. h(x)p(x)    \right|_a^b  - \int_a^b   h(x)  \frac{ d p(x)  }{d x}  dx.
	\end{gather}
	which is actually the same as that obtained through integration by part method given in Eq. (\ref{eqn:NSI_general_qx}). 
	
\end{enumerate}

\section{Discussion}
It can be seen that the integral of Eq. (\ref{eqn:partial_general0}) in  Section \ref{sec:general} can lead to various of integration by part methods by choosing different $\tilde{q}(x)$, which are summarized as follow.

\begin{enumerate}
	
	\item $\tilde{q}(x) = q(x)$: when $q(x)$ is known and the indefinite integral of its inverse is known, it is equivalent to the integration by part method shown in  Eq. (\ref{titlde_qx}).
	\begin{flalign}\label{titlde_qx_disc}
		& I_0 = \left.  {p}(x)  {h}_q(x)  \right|_a^b - I',
	\end{flalign} 
	with
	\begin{gather}
		I' \equiv \int_a^b   {h}_q(x) p'(x) dx; \\
		h_q(x) = \int \frac{1}{q(x)} dx. \nonumber
	\end{gather}
	
	\item $\tilde{q}(x)$: other $\tilde{q}(x)$ close to $q(x)$ can be used if the integral of its inverse $\tilde{h}(x)$ is known and the following integration by part can be used,
	\begin{flalign}\label{titlde_qx_app_disc}
		& I_0 = \left. \tilde{p}(x) \tilde{h}_q(x)  \right|_a^b - \tilde{I}',
	\end{flalign} 
	with
	\begin{gather}
		\tilde{I}' = \int_a^b  \tilde{h}_q(x) \left[\tilde{q}(x) f(x)\right]' dx \\
		\tilde{h}_q(x) = \int \frac{1}{\tilde{q}(x)} dx. \nonumber
	\end{gather}
	
	\item $\tilde{q}(x) = x^{\gamma}$: it is equivalent to the integration by part method shown in  Eq. (\ref{I0_by_part_1}). In particular, when $\gamma=0$ and $h_\gamma(x) =x$, the following is obtained,
	\begin{gather}\label{gamma_disc}
		I_0 = \frac{1}{1-\gamma} \left\{\left. x f(x) \right|_a^b  - I_\gamma'  \right\},
	\end{gather}
	with
	\begin{gather}
		I_\gamma' \equiv \int_a^b  x^{1-\gamma} \left[ x^\gamma f(x) \right]'  dx.
	\end{gather}
	
	In particular, when $\gamma=0$, 
	\begin{gather} \label{gamma0_disc}
		I_0 = \left. x f(x)\right|_a^b  - I_0',
	\end{gather}
	with
	\begin{gather}\label{I0p}
		I_0' \equiv \int_a^b  x f'(x)  dx.
	\end{gather}
	
	In practice, which one of the above integration by part is used depends on the following criteria.
	
	\begin{enumerate}
		\item The analytical formula of $q(x)$ or $\tilde{q}(x)$ and integrals of its inverse $h_q(x)$ or $\tilde{h}_q$ are known.
		
		\item The integral $I'$, $\tilde{I}'$ or $I_\gamma'$ should be non-singular and can be numerically computed efficiently.
		
	\end{enumerate}
	
\end{enumerate}

\subsection{Error Analysis}
Now let's perform the error analysis of Eq. (\ref{I0_all}). Without loss of generality, assuming that $p'(x) =0$ and the following is obtained,
\begin{gather}\label{I0_all_err}
	I_0 = \tilde{I}_b - \left. o(x)   \tilde{h}(x) \tilde{q}(x)  f(x)  \right|_a^b - I_o. \nonumber
\end{gather}
with
\begin{gather}\label{Io}
	I_o \equiv  \int_a^b  f_o(x)  dx, \\
	f_o(x) \equiv o(x) \left[\tilde{h}(x) \tilde{q}(x)  f(x)  \right]', \ \ o(x) = - \ln\left(\frac{\tilde{q}(x) }{q(x)}\right). \nonumber
\end{gather}

It can be seen that  $o(x) \sim 0$ when $\tilde{q}(x) \sim q(x)$. Let's assume $q(x) \sim x^{\gamma}$ and $\tilde{q}(x) \sim x^{\tilde{\gamma}}$, from which the following is obtained,
\begin{gather}
	o(x) = -(\tilde{\gamma} -\gamma) \ln(x) = - \delta_\gamma \ln(x),
\end{gather}
where $\delta_\gamma \equiv \tilde{\gamma} -\gamma$ and $I_o$ of Eq. (\ref{Io}) becomes 
\begin{gather}
	I_o =  \int_a^b  o(x) \left[\tilde{h}(x) \tilde{q}(x)  f(x)  \right]'  dx \\
	= \delta_\gamma \frac{1-\gamma}{1-\tilde{\gamma}} \int_a^b  (-\ln(x)) f(x)  dx \nonumber \\
	\delta_\gamma \frac{ (1-\gamma)}{1-\tilde{\gamma}} p(0) \int_a^b     \frac{-\ln(x)}{x^\gamma} dx = c \mathcal{O}(\delta_\gamma),   \nonumber
\end{gather} 
with 
\begin{gather}
	c =  \frac{  p(0)}{(1-\tilde{\gamma})(1-\gamma)} \left. \left\{  [ 1- (1-\gamma) \ln(x) ] x^{1-\gamma} \right\} \right|_a^b.
\end{gather} 

Compared to the original integral $I_0$, the following relative accuracy is obtained,
\begin{gather}\label{ratio}
	\epsilon = \frac{I_o}{I_0} =  \frac{\delta \gamma}{(1-\tilde{\gamma})}   \frac{\left. \left\{  [ 1- (1-\gamma) \ln(x)] x^{1-\gamma} \right\}  \right|_a^b}{\left. x^{1-\gamma} \right|_a^b}  \\
	\frac{\delta \gamma}{(1-\tilde{\gamma})}   \frac{[ 1- (1-\gamma) \ln(b)] b^{1-\gamma}}{b^{1-\gamma} - a^{1-\gamma}}, \nonumber 
\end{gather}
where $a \ge 0$ has been assumed.

In particular, when $b \le 1$, Eq. (\ref{ratio}) reduces to
\begin{gather}
	\epsilon \le \frac{\delta \gamma}{(1-\tilde{\gamma})}   \frac{1}{b^{1-\gamma} - a^{1-\gamma}} \le \frac{\delta \gamma}{(1-\tilde{\gamma})} \sim \mathcal{O}(\delta \gamma). \nonumber 
\end{gather}

\subsection{The NSI-IBP Algorithm}
Now let's present the numerical singular integral algorithm via the integration by part methods.

For a given singular or nearly singular integral, the following numerical recipe can be used.

\begin{enumerate}
	
	\item If the analytical formulas of $q(x)$ and the integral of its inverse $h(x)$ is known, then the integration by part method of Eq. (\ref{titlde_qx_disc}) can  be used. Otherwise, go to the next step.
	
	\item Find an approximate function $\tilde{q}(x)$ that can closely fits $q(x)$. Also, the analytical formula of the integral of its inverse $\tilde{h}(x)$ should be known. Then the integration by part method of Eq. (\ref{titlde_qx_app_disc}) can be used. Otherwise, go to the next step.
	
	\item Find $\tilde{q}(x) = x^\gamma$ so that the original integrand $f(x)$ can be effectively suppressed at its singularity location, \ie, $x_k^\gamma f(x_k) = 0$. Then, Eq. (\ref{gamma_disc}) can be used. If the integrand of $I_\gamma'$ is non singular and $I_\gamma'$ can be computed numerically, the algorithm stops. Otherwise, go to the next step.
	
	\item Choose $\gamma=0$ and $\tilde{q}(x) = 1$, so that Eq. (\ref{gamma0_disc}) can be used. 
	
	\item Repeat Step 4) until $I_0'$ in Eq. (\ref{I0p}) can be computed numerically, for example, when $\gamma \leq 1$ and is known.
	
\end{enumerate}

\section{Important Classes of Singular Integrals}
Now let's look at the following classes of singular integrals.

\subsection{Singularities of Power-law Decay of $x^{-\gamma}$}
First, let's consider the single singularity point located at $x_k =0$ with $q(x) = x^{\gamma}$. The singular integral of Eq. (\ref{eqn:NSI_general}) reduces to the following,
\begin{flalign}\label{eqn:alpha}
	& I_0 = \int_a^b \frac{p(x)}{x^{\gamma}}  dx.
\end{flalign}

\subsection{Approximate $\tilde{q}(x)$}
When $\gamma$ is not known, it would be wise to choose $\tilde{q}(x) = x^{\tilde{\gamma}}$ with $\tilde{\gamma} \ge \gamma$ and Eq. (\ref{I0_all}) reduces to the following by setting $q(x) = \tilde{q}(x)$,
\begin{gather}\label{I0_power_part}
	I_0 = \left. \tilde{h}(x)  \tilde{q}(x)  f(x)   \right|_a^b 
	- \int_a^b \tilde{h}(x) \left[\tilde{q}(x) f(x) \right]' dx \\
	= \left. \frac{x f(x)  }{1-\tilde{\gamma}}    \right|_a^b 
	- \int_a^b \frac{x^{1-\tilde{\gamma}}}{1-\tilde{\gamma}}  \left[ x^{\tilde{\gamma}-\gamma} p(x) \right]' dx \nonumber \\
	= \left. \frac{x f(x)  }{1-\tilde{\gamma}}    \right|_a^b 
	- \int_a^b \frac{1}{1-\tilde{\gamma}}  \left[  \frac{\tilde{\gamma}-\gamma}{x^\gamma} p(x) +  x^{1-\gamma} p'(x)  \right] dx, \nonumber 
\end{gather}
where Eq. (\ref{hs}) has been used to obtained $\tilde{h}(x)$,
\begin{gather}
	\tilde{h}(x) = \int \frac{1}{\tilde{q}(x)} dx = \frac{x^{1-\tilde{\gamma}}}{1-\tilde{\gamma}} .
\end{gather}

It can be shown in Eq. (\ref{I0_power_part}) that the order of polynomial in the integrand remains the same for arbitrary chosen $\tilde{\gamma}$, which means that the performance of the numerical computation doesn't depend on the choice of $\tilde{\gamma} \neq 1$. 

\subsubsection{Integration by Part Method}
When $\gamma$ is known,  Eq. (\ref{eqn:NSI_general_qx}) reduces to the following,
\begin{gather}\label{I1_part_power}
	I_0  = \left. \frac{x f(x)  }{1-\gamma}    \right|_a^b 
	- \int_a^b \frac{x^{1-\gamma}}{1-\gamma}  p'(x) dx.  
\end{gather} 

\subsubsection{Direct Integration Methods}
The formula for the direction integration methods given in Section \ref{sec:direct} can be shown to be identical to the integration by part method of Eq. (\ref{I1_part_power}).  



\subsection{Logarithmic Singularities of $\ln^{\beta}(x)$}
The singular integral of Eq. (\ref{eqn:NSI_general}) reduces to the following,
\begin{flalign}\label{I0_ln}
	& I_\beta  = \int_a^b  \ln^{\beta}(x) p(x)  dx,
\end{flalign} 
with $q(x) = \ln^{-\beta}(x)$.


\subsubsection{Integration by Part Method}
Eq. (\ref{eqn:NSI_general_qx}) reduces to
\begin{gather}\label{Igamma_part}
	I_\beta = \left. \left[p(x) h_\beta(x) \right]   \right|_a^b  - \int_a^b   p'(x) h_\beta(x) dx,  
\end{gather}
with
\begin{gather}
	h_\beta(x) \equiv \int_{x_0}^x \ln^{\beta}(\chi) d\chi,
\end{gather}
where integral starting point $x_0 \neq 0$ can be chosen arbitrarily and the integral can be obtained iteratively until $\beta \le 0$ as follows,
\begin{gather}
	h_\beta(x) = x \ln^{\beta}(x) - \beta h_{\beta-1}(x), 
\end{gather}
and 
\begin{gather}
	h_0(x) = x; \\
	h_{\beta < 0}(x) = \int_{x_0}^x \frac{1}{\ln^{|\beta|}(\chi)} d\chi.
\end{gather}

In particular, $h_{\beta=1}(x) = x\ln(x) - x$.

\subsubsection{Direct Integration Methods}
Let's consider the case of $h(x) \tilde{q}(x) = x$. By setting  $\tilde{q}(x) = x^{\tilde{\beta}}$, the following is obtained,
\begin{gather}
	h(x) = \frac{x}{x^{\tilde{\beta}}},
\end{gather}
and Eq. (\ref{hsx}) reduces to 
\begin{gather}\label{hsx_ln}  
	I_\beta  = - \beta I_{\beta-1} + \left. x p(x) \ln^{\beta}(x)     \right|_a^b  \\
	- \int_a^b  x \ln^{\beta}(x)  p'(x) dx,  \nonumber 
\end{gather}
which can be computed iteratively until $ \beta  \le 0$,
\begin{gather} \label{Igamma}
	I_{\beta=0}  =  \int_a^b   p(x)  dx;  \\
	I_{\beta}  =  \int_a^b \frac{1}{\ln^{|\beta|}(x)} p(x)  dx. \nonumber
\end{gather}

It is clear that Eq. (\ref{Igamma}) is non-singular and can be computed numerically readily. In particular, when $\beta=0$, 
\begin{gather} \label{Igamma0}
	I_{0}  =  \int_a^b   p(x)  dx.
\end{gather}

Compared to the integration by part method of Eq. (\ref{Igamma_part}), the advantage of this method is that one doesn't need to know the analytical form of $h_\beta(x)$ and simply repeat the numerical integration of Eq. (\ref{hsx_ln}) using $\tilde{q}(x) = x^{\tilde{\beta}}$.

	
	


\subsection{Hybrid Singularities of $x^{-\gamma}\ln^{\beta}(x)$}
The singular integral of Eq. (\ref{eqn:NSI_general}) reduces to the following,
\begin{flalign}\label{I0_ln_hybrid}
	& I_{\gamma,\beta}  = \int_a^b  \frac{\ln^{\beta}(x)}{x^\gamma} p_0(x)  dx.
\end{flalign}  

\subsubsection{Approximate $\tilde{q}(x)$}
Similar to Eq. (\ref{I0_power_part}), when $\gamma$ is not known, Eq. (\ref{I0_all}) reduces to the following by setting $q(x) = \tilde{q}(x)$,
\begin{gather}\label{I0_power_part_hybrid}
	I_{\gamma,\beta}  =  \frac{\left. x f(x)   \right|_a^b - (\tilde{\gamma}-\gamma) I_{\gamma,\beta}  - \beta I_{\gamma,\beta-1}  
		-   I_{\gamma,\beta}'}{1-\tilde{\gamma}}  ,    
\end{gather}
where Eq. (\ref{hs}) has been used to obtained $\tilde{h}(x)$,
\begin{gather}
	I_{\gamma,\beta}' = \int_a^b \frac{x \ln^{\beta}(x)}{x^{ {\gamma}}}  {p}_0'(x)    dx \\
	\tilde{h}(x) = \int \frac{1}{\tilde{q}(x)} dx = \frac{x^{1-\tilde{\gamma}}}{1-\tilde{\gamma}} . \nonumber
\end{gather}

\subsubsection{Integration by Part Method}
Eq. (\ref{eqn:NSI_general_qx}) reduces to
\begin{gather}\label{Igamma_part_hybrid}
	I_{\gamma,\beta} = \left. \left[ {p}(x)  {h}_{\gamma}(x) \right]   \right|_a^b  - \int_a^b    {p}'(x)  {h}_{\gamma}(x) dx \\
	=  \frac{\left. x f(x)   \right|_a^b    - \beta I_{\gamma,\beta-1}  
		-   I_{\gamma,\beta}'}{1- {\gamma}}, \nonumber
\end{gather}
with
\begin{gather}
	{h}_{\gamma}(x) \equiv \frac{x^{1- {\gamma}}}{1- {\gamma}}, \ \  {p}(x) = p_0(x) \ln^{\beta}(x).
\end{gather}

\subsection{Nearly Singular Function of $(x^2+R^2)^{-\gamma}$}
The singular integral of Eq. (\ref{eqn:NSI_general}) reduces to the following,
\begin{flalign}\label{I0_ln_near}
	& I_\gamma = \int \frac{p_0(x)}{(x^2+R^2)^{\gamma}}   dx.
\end{flalign}

\subsubsection{Approximate $\tilde{q}(x)$}
Similar to Eq. (\ref{I0_power_part}), when $\gamma$ is not known, Eq. (\ref{I0_all}) reduces to the following by setting $q(x) = \tilde{q}(x)$,
\begin{gather}\label{I0_power_part_near}
	I_\gamma = \left. \tilde{h}(x)  \tilde{q}(x)  f(x)   \right|_a^b 
	- \int_a^b \tilde{h}(x) \left[\tilde{q}(x) f(x) \right]' dx \\ 
	= - \frac{\tilde{\gamma}}{1-\tilde{\gamma}} I_\gamma  + \left. \tilde{h}(x)  \tilde{q}(x)  f(x)   \right|_a^b    \nonumber \\  
	+ \frac{2\gamma}{1-\tilde{\gamma}} \int_a^b \frac{x^{2} p_0(x)}{(x^2+R^2)^{1+\gamma}} dx  
	- \frac{1}{1-\tilde{\gamma}} \int_a^b \frac{x  p_0'(x)}{(x^2+R^2)^{\gamma}} dx, \nonumber
\end{gather}
where Eq. (\ref{hs}) has been used to obtained $\tilde{h}(x)$,
\begin{gather} 
	\tilde{h}(x) = \int \frac{1}{\tilde{q}(x)} dx = \frac{x^{1-\tilde{\gamma}}}{1-\tilde{\gamma}} .  
\end{gather}

In particular, when $\tilde{\gamma} =$ 0, Eq. (\ref{I0_power_part_near}) reduces to the following, \begin{gather}\label{I0_power_part_near0}
	I_\gamma =     \left. \tilde{h}(x)  \tilde{q}(x)  f(x)   \right|_a^b    \nonumber \\  
	+2\gamma \int_a^b \frac{x^{2} p_0(x)}{(x^2+R^2)^{1+\gamma}} dx  
	-  \int_a^b \frac{x  p_0'(x)}{(x^2+R^2)^{\gamma}} dx.  
\end{gather}

\subsubsection{Integration by Part Method}
Eq. (\ref{eqn:NSI_general_qx}) reduces to
\begin{gather}\label{Igamma_part_near}
	I_{\gamma} = \left. \left[ {p}(x)  {h}_{\gamma}(x) \right]   \right|_a^b  - \int_a^b    {p}'(x)  {h}_{\gamma}(x) dx \\
	=  \left.  \frac{p_0(x)  \ln\left( | (x^2+R^2)^{\frac{1}{2}}+ x| \right) }{(x^2+R^2)^{\gamma-\frac{1}{2}}}   \right|_a^b  
	- I_-  + \left(2\gamma - 1\right) I_+,  \nonumber 
\end{gather}
with
\begin{gather}
	I_- \equiv \int_a^b  \frac{p_0'(x)}{(x^2+R^2)^{\gamma-\frac{1}{2}}} \ln\left( | (x^2+R^2)^{\frac{1}{2}}+ x| \right) dx; \\
	I_+ \equiv  \int_a^b   \frac{xp_0(x)}{(x^2+R^2)^{\gamma+\frac{1}{2}}}  \ln\left( | (x^2+R^2)^{\frac{1}{2}}+ x| \right) dx ; \nonumber \\
	q(x) = (x^2+R^2)^{\frac{1}{2}}; \nonumber \\
	{h}_{\gamma}(x) = \ln\left( | (x^2+R^2)^{\frac{1}{2}}+ x| \right); \nonumber \\
	p(x) =   \frac{p_0(x)}{(x^2+R^2)^{\gamma-\frac{1}{2}}} . \nonumber
\end{gather}

However, both $I_-$ and $I_+$ of Eq. (\ref{Igamma_part_near}) are still nearly singular and another integration by part with $\tilde{q}(x) = 1$ is required.

In particular, when $\gamma = 1/2$, Eq. (\ref{Igamma_part_near}) reduces to
\begin{gather}\label{Igamma_part_near2}
	I_{\gamma} =   \left. p_0(x)  \ln\left( | (x^2+R^2)^{\frac{1}{2}}+ x| \right)  \right|_a^b  \\ 
	- \int_a^b p_0'(x) \ln\left( | (x^2+R^2)^{\frac{1}{2}}+ x| \right) dx. \nonumber
\end{gather}

\subsection{Nearly Singular Function of $(x^2-R^2)^{-\gamma}$}
The singular integral of Eq. (\ref{eqn:NSI_general}) reduces to the following,
\begin{flalign}\label{I0_ln_k}
	& I_\gamma = \int \frac{p_0(x)}{(x^2-R^2)^{\gamma}}   dx.
\end{flalign}

\subsubsection{Approximate $\tilde{q}(x)$}
Similar to Eq. (\ref{I0_power_part}), when $\gamma$ is not known, Eq. (\ref{I0_all}) reduces to the following by setting $q(x) = \tilde{q}(x)$,
\begin{gather}\label{I0_power_part_k}
	I_\gamma = \left. \tilde{h}(x)  \tilde{q}(x)  f(x)   \right|_a^b 
	- \int_a^b \tilde{h}(x) \left[\tilde{q}(x) f(x) \right]' dx \\ 
	= - \frac{\tilde{\gamma}}{1-\tilde{\gamma}} I_\gamma  + \left. \tilde{h}(x)  \tilde{q}(x)  f(x)   \right|_a^b    \nonumber \\  
	+ \frac{2\gamma}{1-\tilde{\gamma}} \int_a^b \frac{x(x-d) p_0(x)}{(x^2-R^2)^{1+\gamma}} dx  
	- \frac{1}{1-\tilde{\gamma}} \int_a^b \frac{ (x-d)  p_0'(x)}{(x^2-R^2)^{\gamma}} dx, \nonumber
\end{gather}
where Eq. (\ref{hs}) has been used to obtained $\tilde{h}(x)$,
\begin{gather} 
	\tilde{q}(x) = (x-d)^{\tilde{\gamma}}, \\
	\tilde{h}(x) = \int \frac{1}{\tilde{q}(x)} dx = \frac{ (x-d)^{1-\tilde{\gamma}}}{1-\tilde{\gamma}} .   \nonumber
\end{gather}

\subsubsection{Integration by Part Method}
Eq. (\ref{eqn:NSI_general_qx}) reduces to
\begin{gather}\label{Igamma_part_k}
	I_{\gamma} = \left. \left[ {p}(x)  {h}_{\gamma}(x) \right]   \right|_a^b  - \int_a^b    {p}'(x)  {h}_{\gamma}(x) dx \\
	=  \left.  \frac{p_0(x)   \ln\left( | (x^2-R^2)^{\frac{1}{2}}+ x| \right) }{(x^2-R^2)^{\gamma - \frac{1}{2}}}   \right|_a^b 
	- I_-  + \left(2\gamma - 1\right) I_+ , \nonumber
\end{gather}
with
\begin{gather} 
	I_- \equiv \int_a^b  \frac{p_0'(x)}{(x^2-R^2)^{\gamma-\frac{1}{2}}} \ln\left( | (x^2-R^2)^{\frac{1}{2}}+ x| \right) dx; \\
	I_+ =   \int_a^b   \frac{xp_0(x)}{(x^2-R^2)^{\gamma+\frac{1}{2}}}  \ln\left( | (x^2-R^2)^{\frac{1}{2}}+ x| \right) dx; \nonumber \\
	q(x) \equiv (x^2-R^2)^{\frac{1}{2}}; \nonumber \\
	{h}_{\gamma}(x) = \ln\left( | (x^2-R^2)^{\frac{1}{2}}+ x| \right); \nonumber \\
	p(x) =   \frac{ (x^2-R^2)^{\frac{1}{2}} }{(x^2-R^2)^{\gamma}} p_0(x) . \nonumber
\end{gather}

However, both $I_-$ and $I_+$ of Eq. (\ref{Igamma_part_k}) are still nearly singular and another integration by part with $\tilde{q}(x) = 1$ is required.

In particular, when $\gamma = 1/2$, Eq. (\ref{Igamma_part_near}) reduces to
\begin{gather}\label{Igamma_part_k2}
	I_{\gamma} =  \left.  p_0(x)   \ln\left( | (x^2-R^2)^{\frac{1}{2}}+ x| \right)  \right|_a^b \nonumber \\ 
	- \int_a^b  p_0'(x) \ln\left( | (x^2-R^2)^{\frac{1}{2}}+ x| \right) dx. \nonumber
\end{gather}

\section{Numerical Examples}

To show the performance of the NSI-IBP method, the following numerical experiments have been carried out. During all numerical computation of the non-singular integrals, the 11-points Gauss-Laguerre  quadrature method with 1001 sampling points is used. Finally, all computation is done in Python 3.12 within the PyCharm 2024.1 (Community Edition) environment.

For all experiments, only singularities at the boundaries are studied, because when the singularities are inside the integration zone, it can be divided into multiple integration zones such that the singularities will be on the boundaries.

Finally, the relative accuracy of the integral computed through the NSI-IBP method, denoted as $\epsilon$, is used to evaluate the performance, which is defined as the relative deviation NSI-IBP value $I$ from its theoretical value $I_{0}$ as follows,
\begin{gather}
	\epsilon \equiv \left|\frac{I - I_{0}}{I_{0}}\right|.
\end{gather}

\begin{figure}[htbp]
	\centering
	\includegraphics[width=1\linewidth]{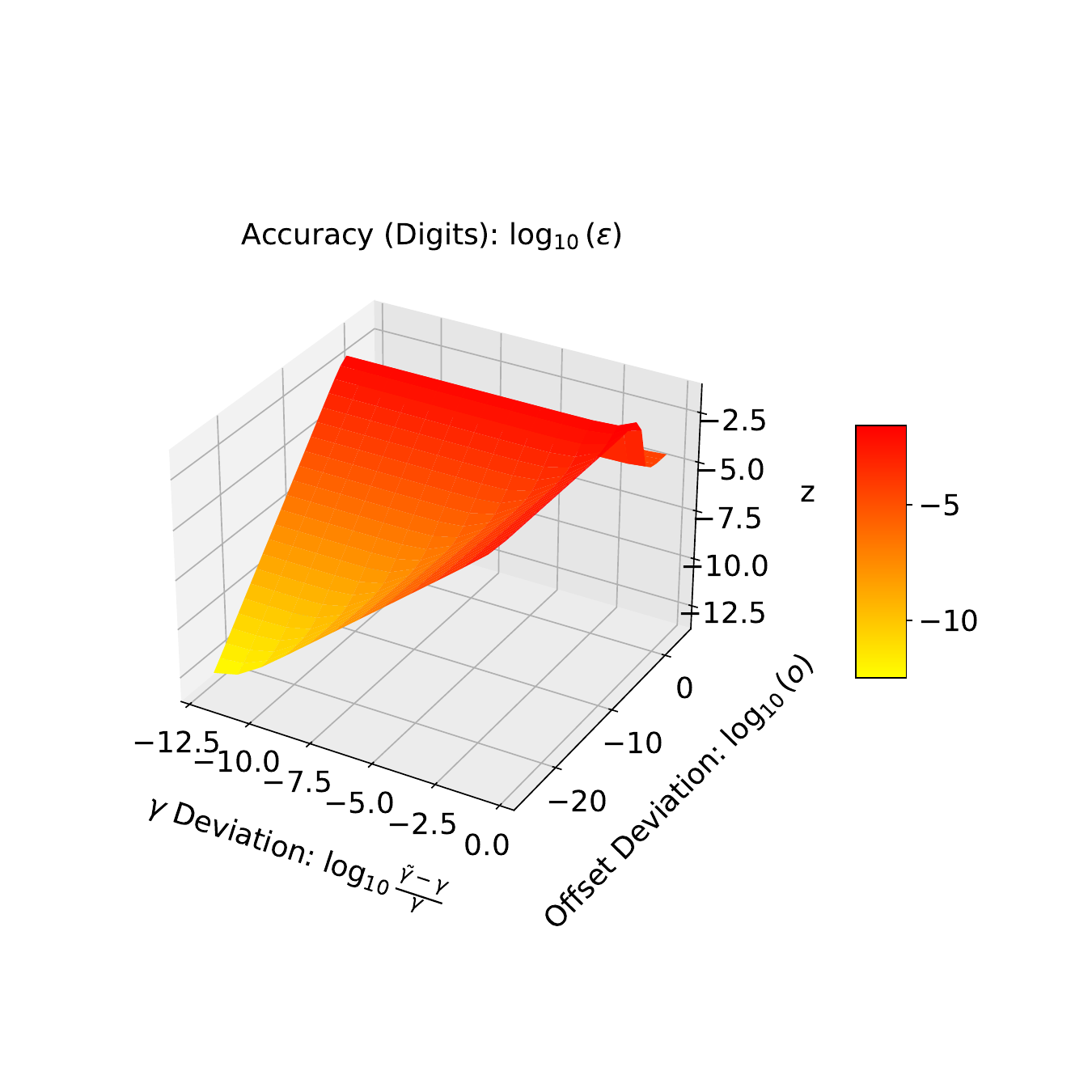}
	\caption{Accuracy surface plot for the power-law singular function of $\gamma = 1/2$, with respect to the deviation of $\Delta \gamma   \in [10^{-12}, 1] \gamma$ for different offsets $o \in [10^{-25}, 10^4]$.}
	\label{surf_gamma_deviations}
\end{figure}

\begin{figure}[htbp]
	\centering
	\includegraphics[width=1\linewidth]{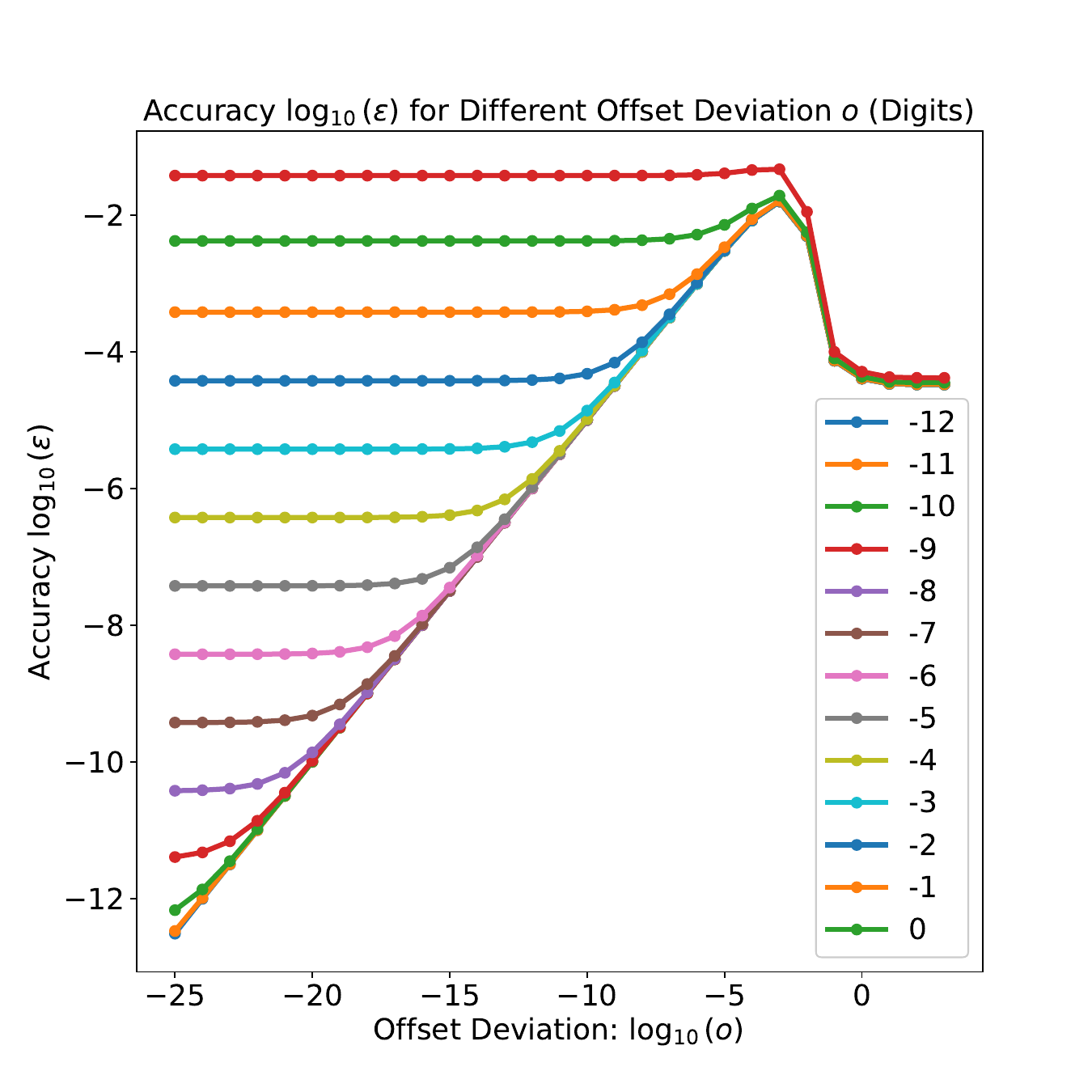}
	\caption{Accuracy lines plot for the power-law singular function with $\gamma = 1/2$, for different offset deviation $o \in [10^{-25}, 10^4]$, when different deviations of $\Delta \gamma   \in [10^{-12}, 1] \gamma$ are used.}
	\label{gamma_deviations}
\end{figure}

\subsection{Singularities of Power-law Decay of $x^{-\gamma}$}
The following integral with power-law decay function is used,
\begin{gather}
	I_{\gamma} = \int_a^{b} \frac{1}{(x+o)^\gamma} dx, \ \  \gamma \in [0, 1),
\end{gather}
with the following theoretical value
\begin{gather}
	I_{\gamma} = \left. \frac{(x+o)^{1-\gamma}}{1-\gamma} \right|_a^b,
\end{gather}
where $\gamma$ and $o \le 0$ are unknown.

The NSI-IBP method of Eq. (\ref{I0_power_part}) has been carried out. Fig. \ref{surf_gamma_deviations} shows the accuracy surface plot of the NSI-IBP method with $\gamma = 1/2$ and $[a, b] = [0, 1]$, when $\tilde{\gamma}$ deviates from its exact value $\gamma$, \ie, $\Delta \gamma \equiv \tilde{\gamma} -\gamma \in [10^{-12}, 1] \gamma$ for different offsets $o \in [10^{-25}, 10^4]$. To get the better view of the accuracy, Fig. \ref{gamma_deviations} shows the line plots, from which the following are observed,

\begin{enumerate}
	\item Smaller deviation, better accuracy: on one hand, it can be seen that the accuracy improves when $\tilde{\gamma}$ reaches its exact value $\gamma$, better then $10^{-13}$ when $\Delta \gamma = 10^{-12}$ for $o = 10^{-25}$. 
	
	\item Large offset, constant accuracy: on the other hand, it can be seen that the accuracy approaches a constant value of $<10^{-4}$ when the offset becomes large $o \gg 1$, which can be understood as: when the offset $o$ becomes large, the integrand is no longer singular and can be expanded into polynomial of some constant order $n$.
	
	\item Worst case: around $o \sim 10^{-3}$, the accuracy is the worst $\epsilon  < 10^{-1.3} $.
\end{enumerate}

\begin{figure}[htbp]
	\centering
	\includegraphics[width=1\linewidth]{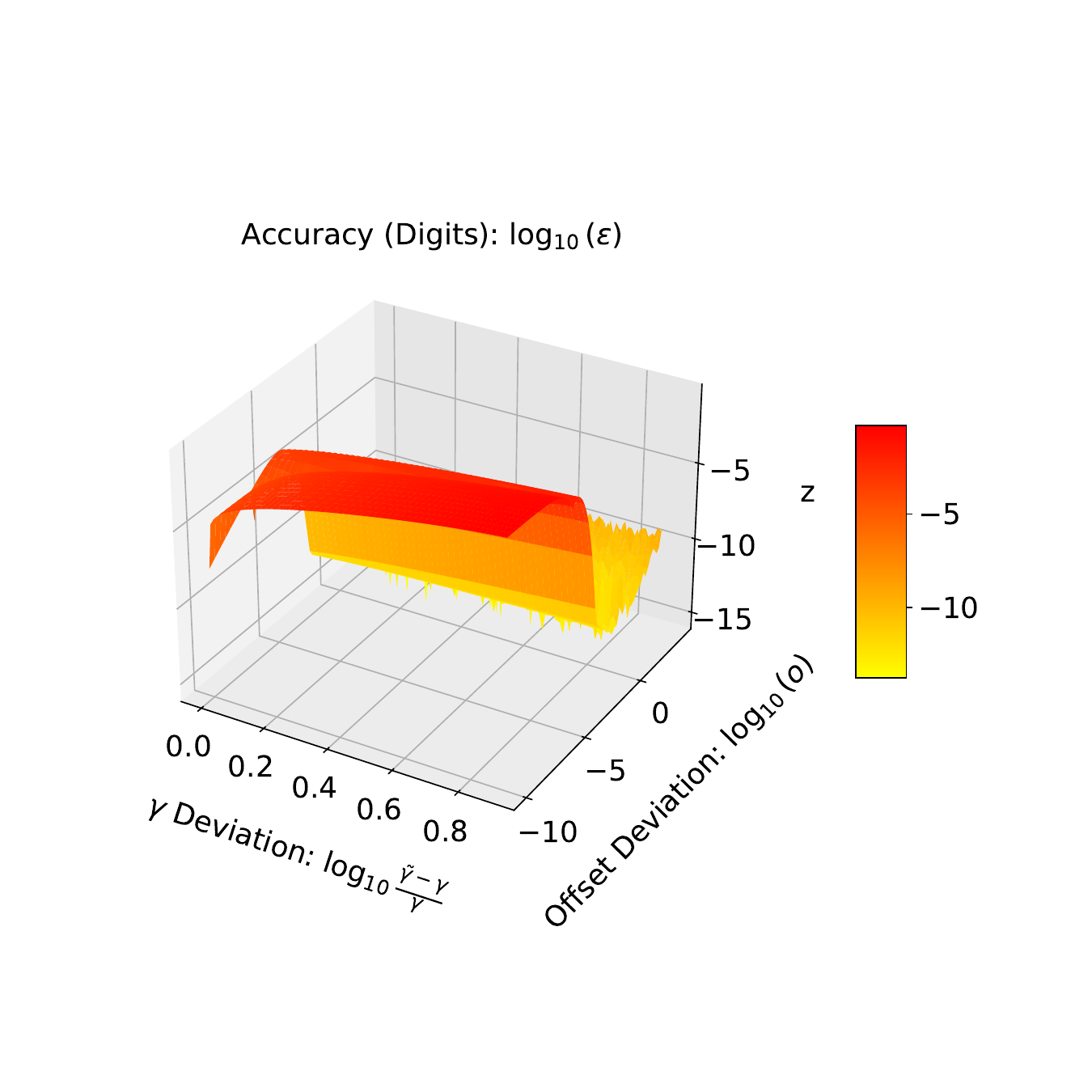}
	\caption{Accuracy surface plot for the hybrid singular function of $\gamma = 1/2$, with respect to the deviations of $\Delta$ for different offsets $o$.}
	\label{surf_gamma_deviations_ln}
\end{figure}

\begin{figure}[htbp]
	\centering
	\includegraphics[width=1\linewidth]{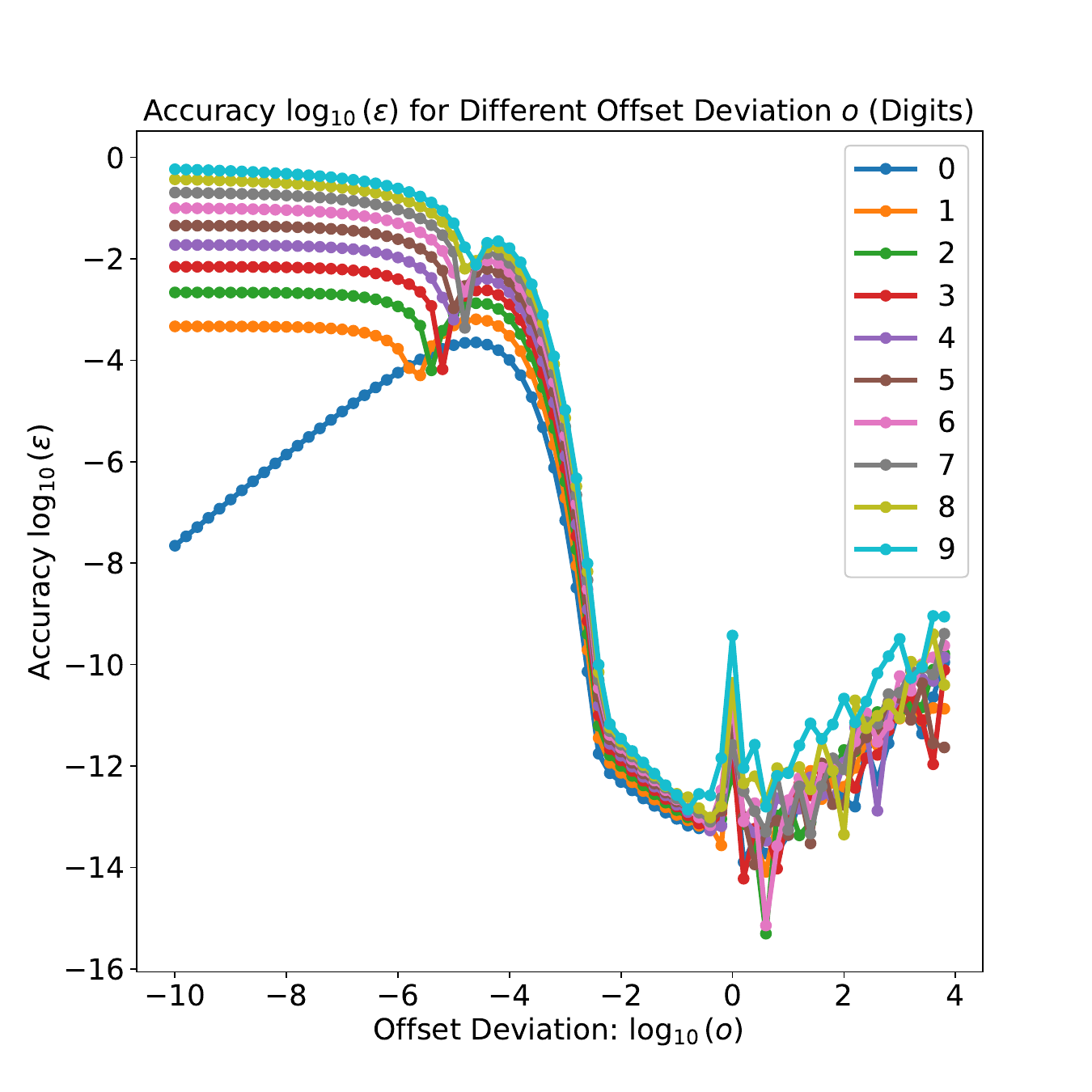}
	\caption{Accuracy lines plot for the hybrid singular function with $\gamma = [0, 9]$, for different offsets $o \in [10^{-25}, 10^4]$, when different deviations $\Delta \gamma   \in [10^{-12}, 1] \gamma$ are used.}
	\label{gamma_deviations_ln}
\end{figure}

\subsection{Hybrid Power-law and Logarithmic Singularities of $x^{-\gamma}\ln(x)$}
The following integral with logarithmic singularity is used,
\begin{gather}
	I_{ln, \gamma} = \int_a^{b}  \frac{\ln(x+o)}{(x+o)^{\gamma}}   dx, \ \  \gamma \in [0, 1),
\end{gather}
with the following theoretical value,
\begin{gather}
	I_{ln, \gamma} = \left. \frac{(1-\gamma)\ln(x)-1}{(1-\gamma)^2} x^{1-\gamma} \right|_a^b,
\end{gather}
where $\gamma$ and $o \le 0$ are unknown.

The NSI-IBP method of Eq. (\ref{I0_power_part_hybrid}) has been carried out with $\tilde{\gamma} =0$. Fig. \ref{surf_gamma_deviations_ln} shows the accuracy surface plot of the NSI-IBP method for $\gamma = 1/2$ within $[a, b] = [0, 0.01]$, when  different $\Delta \gamma$ and offsets  $o$ are used. To get the better view of the accuracy, Fig. \ref{gamma_deviations_ln} shows the line plots, from which the following can be observed, 
\begin{enumerate}
	\item $\gamma$-insensitive accuracy: the accuracy can reach $\epsilon < 10^{-12}$ for all $\gamma \in [0, 9]$.
	
	\item Optimum offsets: the best accuracy occurs around offsets $o \sim [0.1, 10]$, where  $\epsilon < 10^{-15}$ could be achieved for the best case.
\end{enumerate}

\begin{figure}[htbp]
	\centering
	\includegraphics[width=1\linewidth]{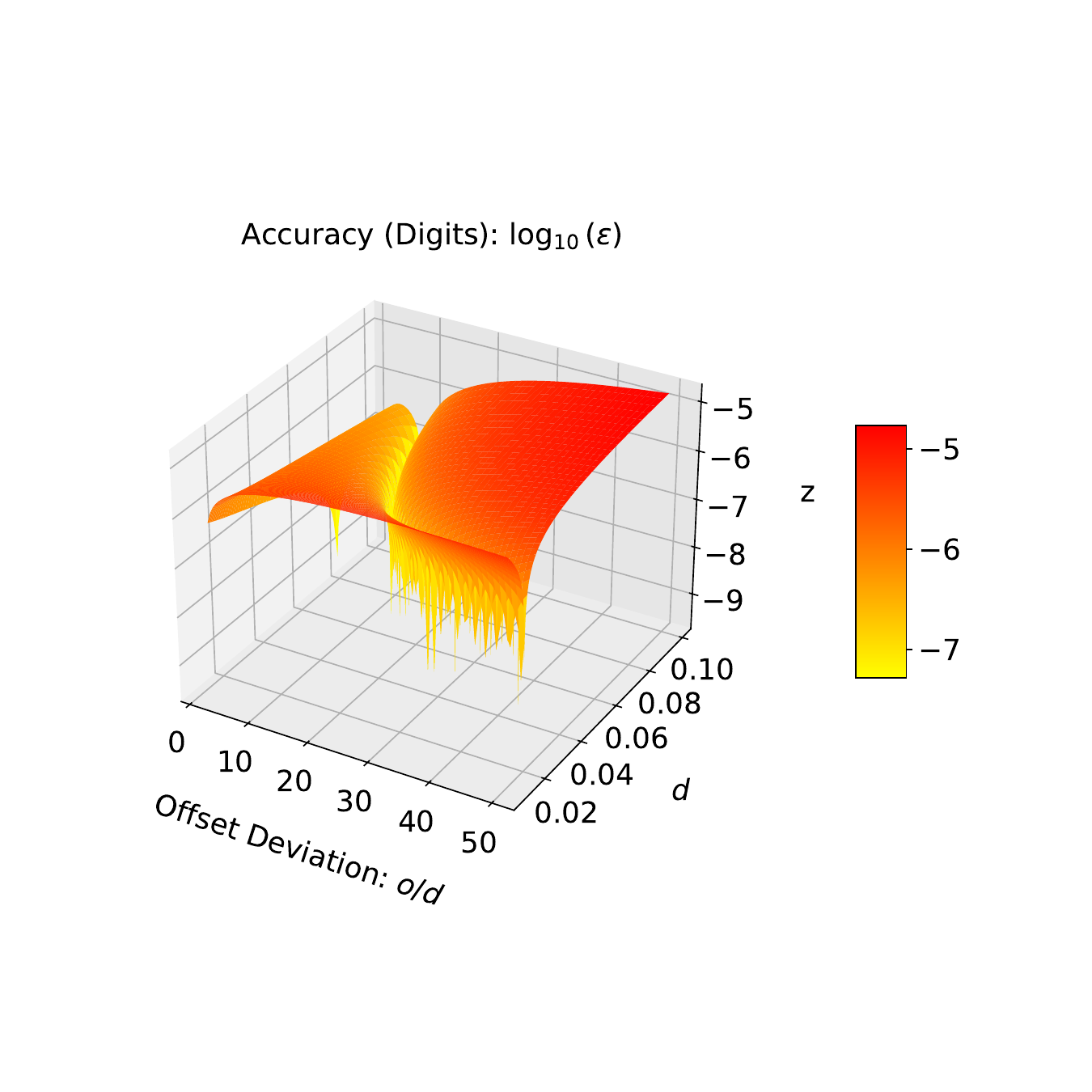}
	\caption{Accuracy surface plot for the electrostatic problem, for different $d \in [0.1, 0.2]$ and offset deviation of $o \in [0, 50] d$ .}
	\label{surf_electrostatic}
\end{figure}

\begin{figure}[htbp]
	\centering
	\includegraphics[width=1\linewidth]{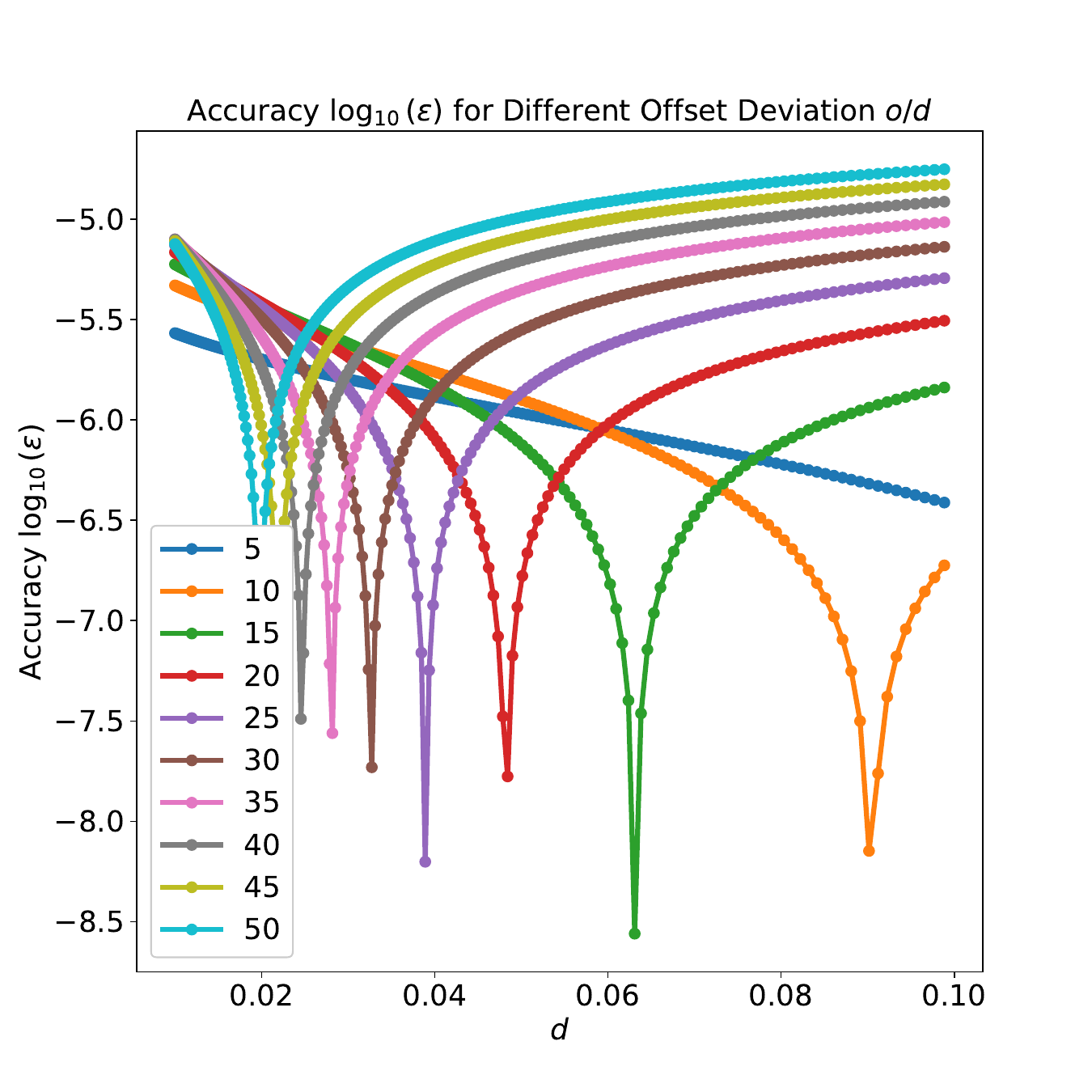}
	\caption{Accuracy lines plot for the electrostatic problem, for different offset deviation of $o \in [0, 50] d$.}
	\label{electrostatic}
\end{figure}

\subsection{Electrostatics Problems}
Now let's consider the electrostatics problem of Eq. (\ref{singular_green_1D}) when $k=0$, which is singular $d=0$ and nearly singular $d \neq 0$,
\begin{gather}\label{Electrostatic}
	I_{\frac{1}{2}}(y) \equiv \int_{a(y)}^{b(y)}  \frac{1}{ \left( x^2 + R^2 \right)^{\frac{1}{2}} }   dx, 
\end{gather}
with the following theoretical value,
\begin{gather}
	I_{\frac{1}{2}}(y) = \left. \ln\left( \sqrt{x^2 + R^2} +x \right) \right|_{a(y)}^{b(y)},
\end{gather}
where $d=\sqrt{x^2 + y^2}$ and the following has been used
\begin{gather}
	\overline{p}_0(x, y) = \frac{\overline{J}(x'-x, y'-y)}{4 \pi }  =1.
\end{gather}

The NSI-IBP method of Eq. (\ref{I0_power_part_near}) is used with $\tilde{\gamma} = 1$. Fig. \ref{surf_electrostatic}  shows the accuracy surface plot of the NSI-IBP method for $d \in [0.1, 0.2]$ within $[a, b] = [0, d]$, when different offsets $o \in [0, 50] d$ are used. To get the better view of the accuracy, Fig. \ref{electrostatic} shows the line plots, from which the following can be observed, 
\begin{enumerate}
	\item $\epsilon<10^{-4}$: the accuracy is smaller than $10^{-4}$ for all combinations of $(d, o)$.
	
	\item Optimum offsets: the best accuracy can be obtained for each $d$ by choosing appropriate offset $o$.
\end{enumerate}

\begin{figure}[htbp]
	\centering
	\includegraphics[width=1\linewidth]{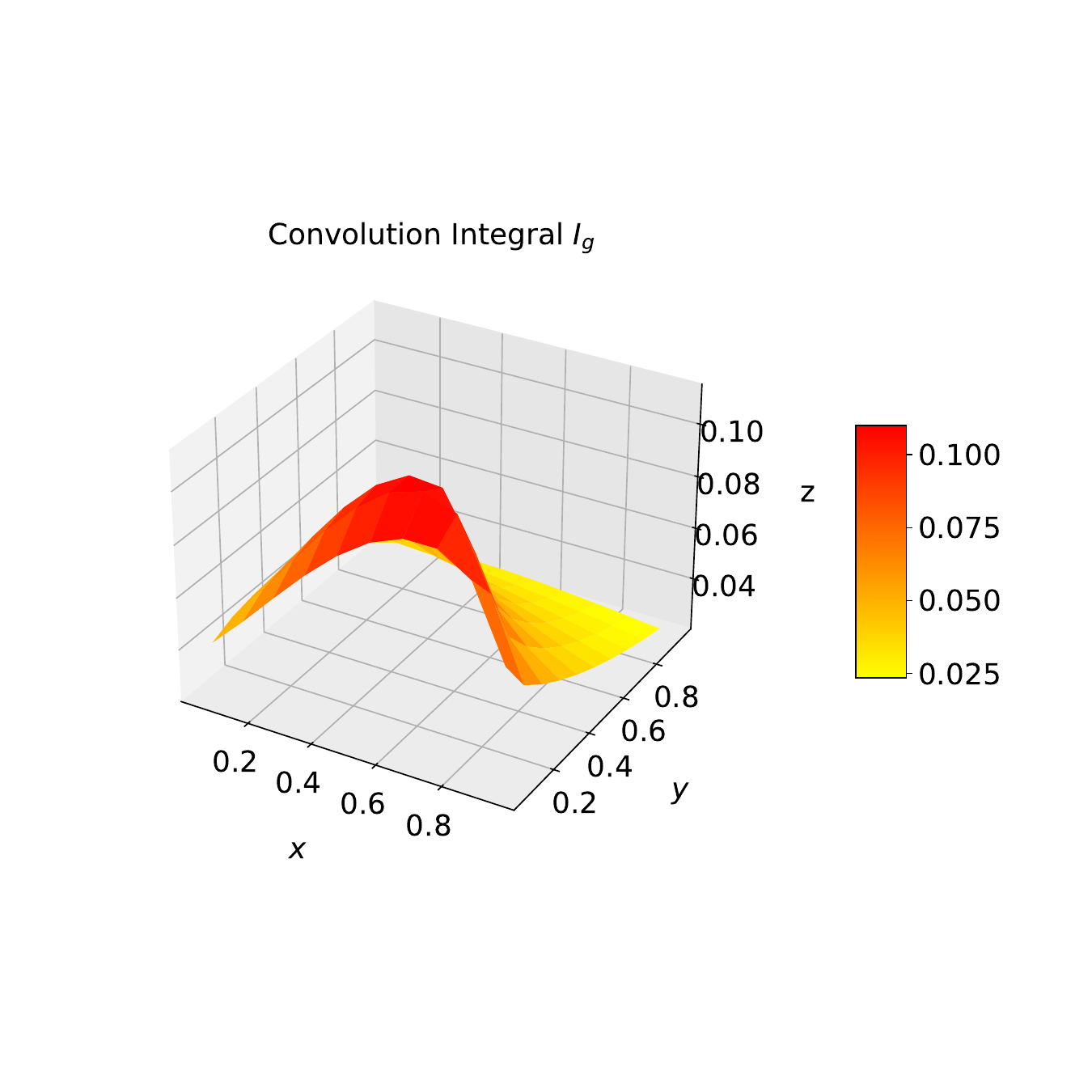}
	\caption{Surface plot of the convolution integral of the scalar Green's function with the RWG current basis  (only on positive triangle).}
	\label{fig_electromagnetic_I}
\end{figure}

\begin{figure}[htbp]
	\centering
	\includegraphics[width=1\linewidth]{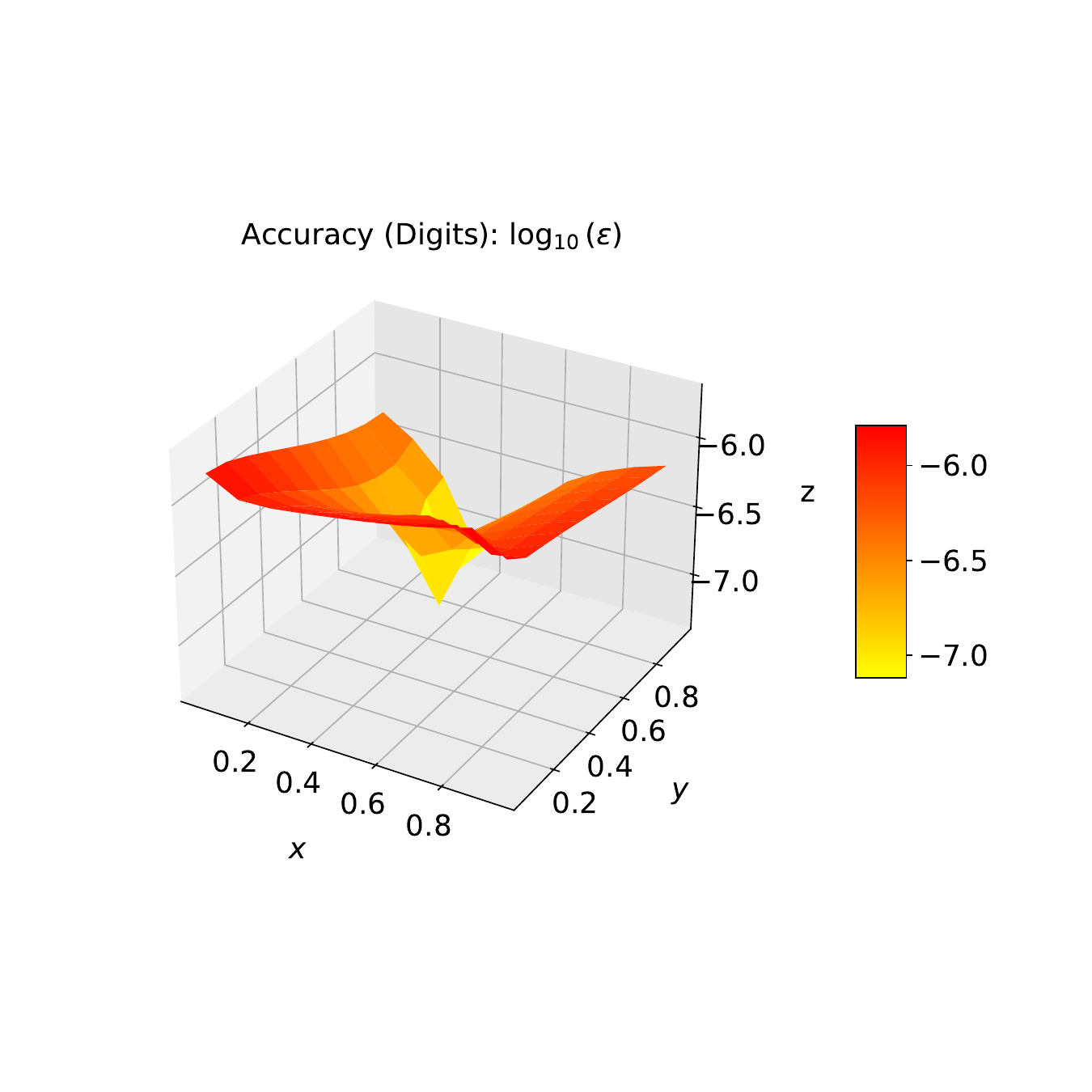}
	\caption{Relative accuracy of the convolution integral of the scalar Green's function with the RWG current basis  (only on positive triangle).}
	\label{fig_electromagnetic}
\end{figure}

\subsection{Electromagnetics Problems}
Let's consider the singular or nearly singular electromagnetics convolution integral $I_g(\overline{r}') $ of Eq.(\ref{singular_green}) of the scalar Green's function with the surface current defined on a RWG triangle pair \cite{Rao_1982} and the observation point $(x', y', z')$ is set to the same RWG triangle pair, as shown in Fig. \ref{fig:problem} , 
\begin{gather}\label{Electromagnetics}
	I_g(\overline{r}') = \int_{0}^1  \int_{0}^{1-y}  \frac{ \overline{p}_0(x, y)}{ \left( x^2 + R_0^2 \right)^{\frac{1}{2}} }   dx dy, 
\end{gather}
where $R_0=\sqrt{x^2 + y^2+(z'-z)^2}$ and the following has been defined,
\begin{gather}
	\overline{p}_0(x, y) = \frac{e^{-j k \left[ x^2 + R_0^2 \right]^{\frac{1}{2}}  }}{4 \pi }  \overline{J}_{RWG}(x'-x, y'-y).
\end{gather} 
with the RWG current basis given by
\begin{gather}
	\overline{J}_{RWG}(x, y) = \left\{ \begin{matrix}
		\frac{\ell}{2A_+} \left( {\bf r}_+  -{\bf v}_+ \right)	\\
		\frac{\ell}{2A_-} \left( {\bf v}_- - {\bf r}_-  \right)
	\end{matrix} \right.,
\end{gather}
where $\ell$ is the length of the common edge of the RWG triangle pair and $+/-$ denote parameters defined on the positive/ negative triangles respectively.

The NSI of Eq. (\ref{Igamma_part_near2}) has been used to compute the convolution integral of Eq. (\ref{Electromagnetics}) with $k=1$, for the RWG basis current of $J_{x, RWG}$ and on the positive triangle. Without loss of generality, the RWG triangle pair is chosen to form a square of unit length, \ie, ${\bf v}_+ = [0, 0]$ and ${\bf v}_- = [1, 1]$. The observation points are $(x, y)  \in [0.05:0.1:0.95]$, as shown in Fig. \ref{fig:problem} .  Also, direct integral with 100,000 sampling points is used as the ground truth for comparison.  Fig. \ref{fig_electromagnetic_I} shows the convolution integral surface plot. Finally, Fig. \ref{fig_electromagnetic} shows the accuracy surface plot, from which it can be seen that the relative accuracy is well below $10^{-6}$.

\section{Conclusion}\label{sec:con}
In this paper, a general framework to compute the singular or nearly singular integrals via the integration by part, or NSI-IBP method, has been presented. A general integration by part formula has been shown to be able to recover all well-known IBP methods. The NSI-IBP method doesn't even need to know the exact forms of the singular or nearly singular integrands. By choosing appropriate singular or nearly singular function with known analytical integral, the singular and nearly singular integral can be transformed to non-singular integral, plus the boundary values. To make the NSI-IBP method more computation friendly, appropriate singular or nearly singular function with known analytical integral has to be chosen. The thumb of rule is to find the approximate function with known analytical integral that best fits the singular integrand. Numerical experiments have been carried out on various singular  integrals such as the power-law decaying function, the logarithmic function and their hybrid products, which shows that the relative accuracy can be as good as $10^{-15}$ when the exact singular function is known and still works well even when the exact singular function is not known. What's more, nearly singular integrals are numerically evaluated for both the electrostatic applications, as well as Computational Electromagnetics (CEM) involving the singular scalar Green's function on the RWG triangle pair, which shows that the NSI-IBP method is very promising in many important applications involving singular integrands, even its exact form is not known. Finally, although 1D scenario is used to present the method, higher-dimensional singular integrals can be dealt with the NSI-IBP method because they can be reduced to many 1D integrals.

\newpage

%
%
%
%

\vfill

\end{document}